\documentclass[11pt]{article}
\usepackage{graphicx}
\pdfoutput=1
\usepackage{geometry}    
\usepackage{authblk}
\usepackage{amssymb}
\usepackage{amsmath}
\usepackage{epstopdf}
\usepackage{amscd}

\usepackage[utf8]{inputenc}
\usepackage{caption}
\usepackage{subcaption}

\usepackage[linktocpage=true]{hyperref}
\hypersetup{
colorlinks=true,
citecolor=blue,
linkcolor=blue,
urlcolor=blue}

\DeclareFontFamily{OT1}{pzc}{}
\DeclareFontShape{OT1}{pzc}{m}{it}{<-> s * [1.10] pzcmi7t}{}
\DeclareMathAlphabet{\mathpzc}{OT1}{pzc}{m}{it}

\def\pd{\partial}

\def\zt{\tilde{z}}

\usepackage{setspace}
\begin{document}

\title{A Pedagogical Introduction to Holographic Hadrons}
\author{Sophia~K.~Domokos$^{1*}$,~ Robert~Bell$^{2*}$,~ Trinh~La$^{3*}$,~ Patrick~Mazza$^4$\footnote{{$^1$ Dept. of Physics, $^2$Dept. of Mathematics, $^3$Col. of Engineering and Computing Sciences, $^4$Dept. of Electrical and Computer Engineering, New York Institute of Technology, 16 W. 61st Street, New York, NY 10023. Email: sdomokos@nyit.edu. }}}

\maketitle

\abstract{ String theory's holographic QCD duality makes predictions for hadron physics by building models that live in five-dimensional (5D) curved space. In this pedagogical note, we explain how finding the hadron mass spectrum in these models amounts to finding the eigenvalues of a time-independent, one-dimensional Schr\"{o}dinger equation. Changing the structure of the 5D curved space is equivalent to altering the potential in the Schr\"{o}dinger equation, which in turn alters the hadron spectrum. We illustrate this concept with three holographic QCD models possessing exact analogs in basic quantum mechanics: the free particle, the infinite square well, and the harmonic oscillator. In addition to making aspects of holographic QCD accessible to undergraduates, this formulation can provide students with  intuition for the meaning of curved space. This paper is intended primarily for researchers interested in involving early-stage undergraduates in research, but is also a suitable introduction to elements of holographic QCD for advanced undergraduate- and beginning graduate students with some knowledge of general relativity and classical field theory.}
\newpage



\section{Introduction}

We have known for over fifty years that protons are made up of quarks and gluons. We have pinned down the masses and couplings of quarks to a startling degree of accuracy. Yet we still don't know why the proton's mass is almost exactly a factor of 100 greater than the sum of the masses of its constituent quarks. This mystery persists because quarks and gluons are {\em strongly coupled} at low energies: they interact so forcefully and often that our usual calculational tools -- based almost entirely on perturbation theory -- fail.

Holographic duality, or  ``holographic QCD" (hQCD) when applied to hadrons \cite{Erlich:2005qh, Karch:2006pv, Sakai:2004cn}, emerged in the early 2000s as a new way to tackle strongly coupled systems. hQCD in particular posits that strongly coupled quarks and gluons in our (3+1)-dimensions (4D) are equivalent  -- or ``dual"-- to a weakly coupled theory living in a  curved (4+1)-dimensional (5D) spacetime.\footnote{The statement of this ``holographic duality" is actually even more powerful: it is really a map between QCD at any energy and a string theory in curved space. For our purposes, however, this low energy limit is sufficient.} This means one could use the classical 5D system to make predictions about the 4D theory's strongly-coupled physics.  Indeed, the classical fields in the 5D spacetime  correspond to the degrees of freedom we see at low energies in 4D: protons, pions, and other quark-gluon composites we collectively call ``hadrons". The 5D fields' properties, meanwhile, translate to predictions for the properties of these hadrons (as detailed further below).

This approach has proved quite successful: even rudimentary hQCD models reproduce hadron spectra and couplings to within about $\sim$15\% of measured values \cite{Erlich:2005qh, Karch:2006pv, Sakai:2004cn}. 

 \begin{figure}
     \centering
         \includegraphics[width=10cm]{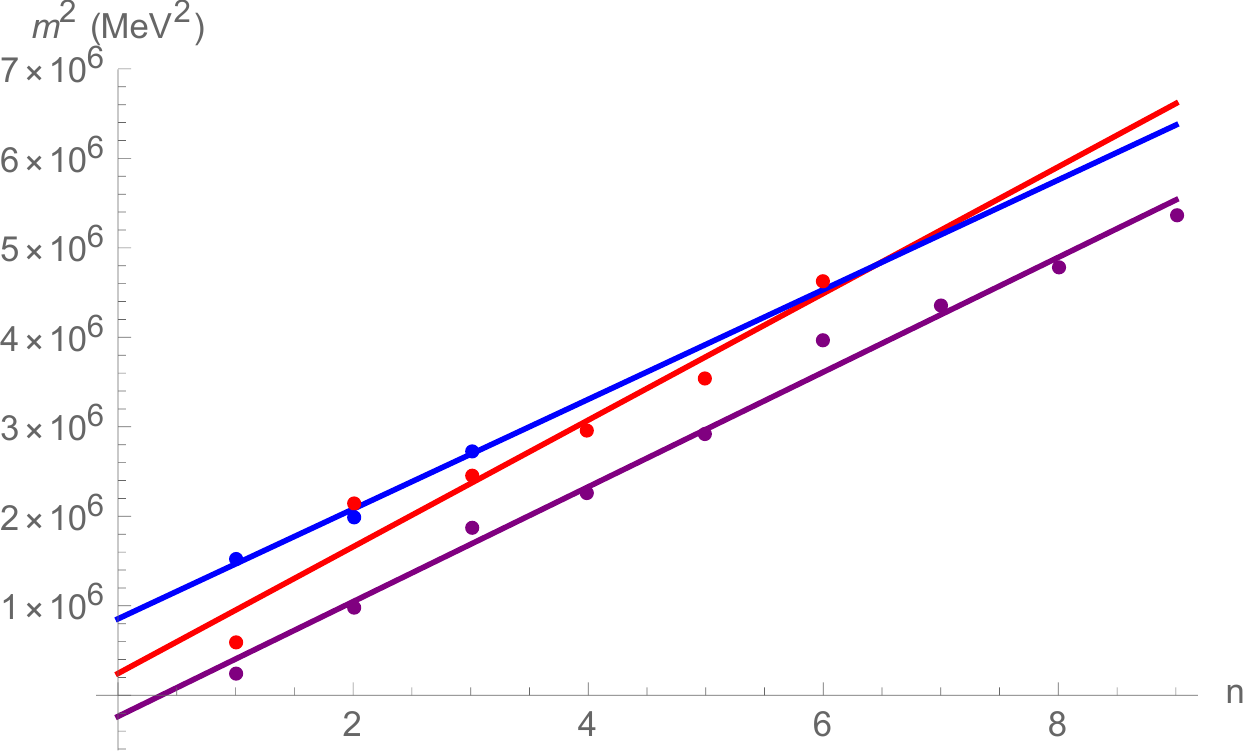}
        \caption{Measured values \cite{Zyla:2020zbs} for meson mass-squared ($m^2$) vs excitation number ($n$) for $\rho$, $a_1$, and $f_0$-type mesons. Other mesons and baryons display similar behavior.}\label{fig:MesonMasses}
\end{figure}

Predictions by hQCD models depend crucially on the shape of the 5D spacetime -- that is, on its metric. The metric should, first of all, be chosen to reproduce the (myriad) patterns observed in the hadron spectrum. We focus here on the patterns highlighted in Figure  \ref{fig:MesonMasses}: the hadron spectrum is discrete; there are several copies of each hadron, which have all of the same quantum numbers but heavier masses; and, the masses-squared of these copies increase roughly linearly. The metric must also have a specific form (``anti-de Sitter") near its boundary, to guarantee the existence of a holographic map in the first place -- that is, that there indeed exists a 4D system the 5D spacetime corresponds to.

In this work, we show that  choosing a 5D metric satisfying these criteria amounts to choosing the right potential in a time-independent Schr\"{o}dinger equation. 

We first transform the equation of motion that determines the spectrum into Schr\"{o}dinger form for arbitrary metric. We then examine the spectrum and Schr\"{o}dinger potential in three examples of hQCD models: (1) Pure anti de-Sitter space (AdS), which admits holographic duality but produces a continuous hadron spectrum because its potential asymptotes to that of a free particle.
(2) AdS cut off at a finite radial value (as in \cite{Erlich:2005qh}) to produce an infinite square well potential.  Here the spectrum is discrete, but mass-squared increases quadratically. (3) A metric like that of \cite{Karch:2006pv}, which yields a harmonic oscillator potential, and producing a discrete spectrum with the desired linear increase of mass-squared.

\subsection{Notation and Conventions}

We use (1) Einstein summation notation, in which all repeated indices are summed over; (2) a ``mostly plus" metric; and (3) ``natural" units in which $\hbar=c=1$. With these conventions, the infinitesimal distance measure in flat 4D (``Minkowski") spacetime is given by
\begin{align}
ds^2 = -dt^2 + (dx^1)^2 + (dx^2)^2+ (dx^3)^2  = \sum\limits_{\mu,\nu =0}^3 \eta_{\mu\nu}dx^\mu dx^\nu := \eta_{\mu\nu}dx^\mu dx^\nu~,
\end{align}
where $\eta_{\mu\nu}$ is the Minkowski metric,  time is parameterized by $t:= x^0$ and the three space dimensions by $x^1$, $x^2$, $x^3$.

In natural units, 
the time-independent Schr\"{o}dinger equation for a particle of mass $M$ and potential $\tilde{V}$ along a single spatial dimensions parameterized by $z$ becomes
\begin{align}
-\frac{\hbar^2}{2M} \Psi''(z) + \tilde{V}(z)\Psi(z) = \tilde{E}\Psi(z)\qquad\Rightarrow \qquad \Psi'' - V(z)\Psi = - m^2\Psi~,
\end{align}
where we  redefine the potential and the eigenvalue to eliminate the constant factors of mass as  $V(z):=2M\tilde{V}(z)$, $m^2:=2M\tilde{E}$. This is the form of the Schr\"{o}dinger equation we use below.

\section{Generic Holographic Models In Schr\"{o}dinger Form}\label{generalmetric}
\subsection{The 5D Metric and The Asymptotically AdS Condition}
In this section, we describing a generic version of a hQCD model, and explain why the metric must be anti de Sitter near its boundary. Let's say the 5D metric is
\begin{align}\label{generalmetric}
d^2x = g_{MN}dx^M dx^N = A(z)^2(\eta_{\mu\nu} dx^\mu dx^\nu + dz^2)~,\qquad\text{where}\quad z\in (0,\infty)~,
\end{align}
for a generic function $A(z)$.
Here capitalized Latin indices $M,N = 0,1,2,3,4$ run over all five spacetime coordinates, and lower case Greek indices $\mu,\nu=0,1,2,3$ run over the four Minkowski dimensions.  This metric preserves the 4D Lorentz symmetry of the spacetime inhabited by the strongly coupled quarks and gluons (parameterized by $x^\mu$). The fifth dimension is parameterized by $z$, with the boundary of the space lying at $z=0$. 

The structure of this space gives some intuition for the holographic correspondence between 4D and 5D: one can think of the 4D Minkowski directions in \eqref{generalmetric} as housing the 4D hadrons; each hadron has a ``wavefunction"  in the extra  dimension ($z$), which encodes its characteristics and behavior. This is a bit of a an oversimplification: the 4D and 5D systems exist independently of each other, and are connected by a map translating one to the other.

Still, the essence of hQCD is that the 4D and 5D theories are just different descriptions of the {\it same} physical system. This means that physical degrees of freedom (e.g. particle states) can be mapped onto each other one-to-one. Both theories must thus contain the same {\em number} of physical degrees of freedom -- and by matching degrees of freedom between the two theories, we can understand why the 5D metric must be anti-de Sitter (AdS) near its boundary. 

The argument is as follows \cite{Baggioli:2019rrs}: The number of degrees of freedom in the 4D quark-and-gluon theory is proportional its volume (it is an extensive quantity). Taking 3D space to be a cube of side $L_4$, discretized into a grid with lattice spacing $\epsilon$, the number of degrees of freedom is then proportional to the number of lattice points, 
\begin{align}\label{N4D}
N_{\rm{4D}}\propto (L_4/\epsilon)^3~.
\end{align}
This counting works differently for gravitating systems, where the number of degrees of freedom is  proportional to the {\em area of the space's boundary} \cite{tHooft:1993dmi, Susskind:1994vu}.
Because our 5D theory lives in curved spacetime, it must have gravity. Again using $\epsilon$ to define the shortest length-scale in the system, we regularize the location of the boundary at $z=0$ in the metric \eqref{generalmetric} to $z=\epsilon$. The degrees of freedom living in the 5D spacetime is
\begin{align}\label{N5D}
N_{\rm{5D}} \propto \int_{z=\epsilon} d^3x \sqrt{-g_\pd} = A(\epsilon)^3 \int d^3x = A(\epsilon)^3 L_4^3
\end{align}
where 
${g_\pd}$ is determinant of the metric restricted to the boundary.  This factor ensures that the volume is independent of coordinate choice:  $\sqrt{-g_\pd}$ cancels out the Jacobian from coordinate transformations of the volume $d^3x$. For $N_{\rm{4D}}$ and $N_{\rm{5D}}$ to scale identically with $\epsilon$, we must have $A(z)\sim 1/z$. The metric of anti de Sitter space (AdS) is simply \eqref{generalmetric} with $A=L/z$ for some constant $L$. The latter thus amounts to requiring that the metric be AdS near $z=0$.
(See e.g. Baggioli's nice introduction to applied holography \cite{Baggioli:2019rrs} for more details.)

\subsection{Action and Reduction to Schr\"{o}dinger Equation}
Having established the basic structure of the 5D metric, we now introduce the classical fields  that correspond to hadrons in 4D. The fields' helicities and (appropriately defined) charge- and parity-conjugation properties determine which hadrons they correspond to. For example, vector fields correspond to vector mesons. As we are interested only in the heuristic scaling of the spectrum, we restrict ourselves to scalar fields, dual to spin 0 hadrons.  
The fields' behavior is determined by an action on the 5D spacetime, the simplest form of which is
\begin{align}
S_{\rm{5d scalar}} = \int d^4x dz \sqrt{-g} \left( \frac{1}{2} \pd_M\Phi\pd_N\Phi  g^{MN}\right) ~,
\end{align}
where $g$ is the determinant of the 5D metric and $\pd_M$ is a partial derivative with respect to $x^M$.  This is identical to the action of a massless scalar field in a flat space, except for the  $\sqrt{-g}$ and the (inverse) metric $g^{MN}$, which as above make the action invariant under coordinate transformations. Each factor of the metric $g_{MN}$ contributes $A(z)^2$ and each inverse metric $g^{MN}$ contributes $A(z)^{-2}$, so the action becomes
\begin{align}\label{generalScalarSwithAz}
S_{\rm{scalar}}=\int d^4x dz A(z)^3 \left( \frac{1}{2} \pd_\mu\Phi\pd_\nu\Phi  \eta^{\mu\nu} +  \frac{1}{2}\pd_z\Phi \pd_z\Phi   \right)~.
\end{align}
with Euler-Lagrange equation
\begin{align}
 A(z)^3 \eta^{\mu\nu} \pd_\mu \pd_\nu \Phi  + \pd_z ( A(z)^3 \pd_z \Phi )= 1~.
\end{align}
We now use separation of variables and the ansatz which corresponds to a state of constant energy $E$ and three-momentum $\vec{p}$:
\begin{align}
\Phi(x,z) = e^{-iEt}e^{i\vec{p}\cdot\vec{x}}\varphi(z)~.
\end{align}
This gives
\begin{align}\label{genEQ}
(E^2-\vec{p}^2)A(z)^3\varphi(z) + \pd_z A(z)^3\pd_z \varphi(z) :=m^2A(z)^3\varphi(z) + \pd_z A(z)^3\pd_z \varphi(z)  =0~.
\end{align}
The 4D mass-energy relation, $(E^2-\vec{p}^2)\equiv m^2$ appears as an eigenvalue in the final equation. $m$ is thus the mass of the 4D hadron, which leads to the requirement that $m^2>0$ for physical particle states.

We can transform \eqref{genEQ} into a time-independent Schr\"{o}dinger equation by defining $\varphi(z) = B(z)\psi(z)$, plugging into \eqref{genEQ},
\begin{align}
\frac{\pd_z (A^3 B')}{A^3B}\psi + \frac{2A^3B'+\pd_z(A^3)B}{A^3B}\psi' +\psi'' = -m^2\psi~,
\end{align}
and choosing  $B(z)$ such that the $\psi'$ terms vanish:
\begin{align}
2B'A^3+\pd_z(A^3)B=0 \qquad\Rightarrow\qquad B=A^{-3/2}~.
\end{align}
The potential is thus
\begin{align}\label{scalarSchro}
V(z) = \frac{1}{4} A^{-6}\left[ 2A^3\pd_z^2(A^3)-( \pd_zA^3 )^2\right]~.
\end{align}
{\bf Boundary Conditions:} Solutions to this Schr\"{o}dinger equation -- dual to hadrons of finite mass and zero spin -- must also satisfy boundary conditions. In this case, these are such that the action \eqref{generalScalarSwithAz} be finite for a given solution. This is equivalent to the normalizability of the wave function in quantum mechanics, and yields the requirement
\begin{align}
\int dz A(z)^3\varphi^2 = \int dz \psi^2 =\text{finite}~,
\end{align}
or $\psi(z\rightarrow0)\sim z^\alpha$ for $\alpha>- 1/2$ and $\psi(z\rightarrow\infty)\sim z^\beta$ where $\beta < -1/2$. As usual, this equation can only be satisfied for special values of the eigenvalue $m^2$, which determines the mass spectrum of the scalar particles. 

\section{Holographic QCD in Three Examples}
We now build up an hQCD model starting with a simple metric, which we modify bit by bit until we achieve the desired behavior of the experimental hadron spectrum.

\subsection{Anti de Sitter space}
\begin{figure}
     \begin{center}
         \includegraphics[width=10cm]{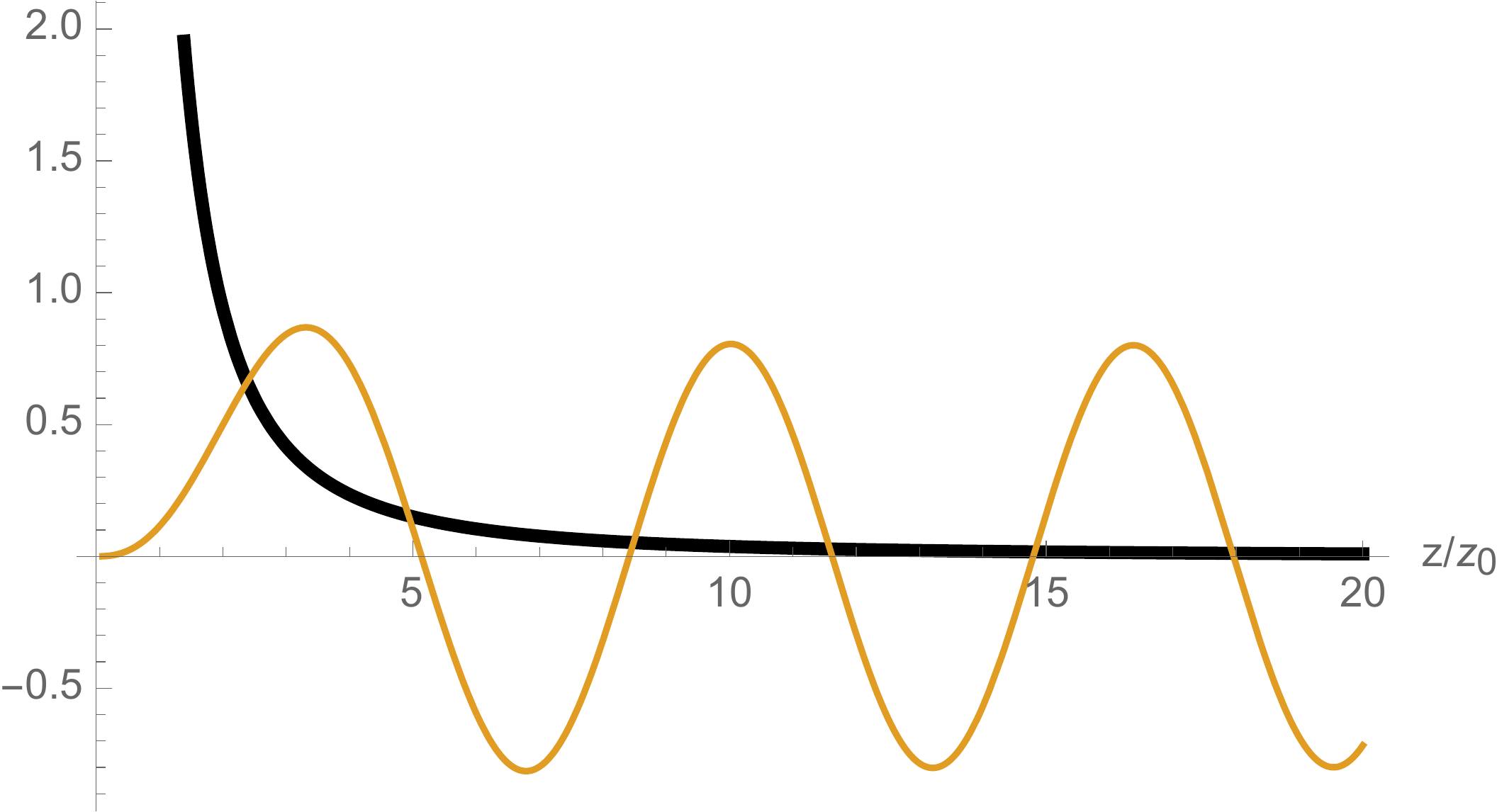}
     \end{center}
        \caption{The potential (black) and a sample wavefunction (orange) for pure anti- de Sitter space (AdS) are shown as a function of a dimensionless coordinate $z/z_0$ (for some scale $z_0$). Note that the potential dies off as $z\rightarrow\infty$, leaving a free particle with the standard sinusoidal wavefunction.}         \label{fig:AdS}
\end{figure}

The simplest example of a metric that is AdS near its boundary is simply AdS all the way through: $A(z)=L/z$ for some scale $L$. Using \eqref{scalarSchro}, the Schr\"{o}dinger potential for pure AdS is
\begin{align}
V_{\text{AdS}} = \frac{15}{4z^2}~,
\end{align}
shown in Figure \ref{fig:AdS}. The arbitrary scale $L$ has dropped out. $V_{\text{AdS}}$ repels particles from the boundary at $z=0$ toward $z\rightarrow\infty$, where it dies off quickly to leave a free particle potential. ($V_{\text{AdS}}$ is in fact identical to the centrifugal potential of the radial wave equation in 3D, which has a similar effect.) The Schr\"{o}dinger equation
\begin{align}
\psi''(z)-\frac{15}{4z^2}\psi = -m^2\psi
\end{align}
has solution
\begin{align}\label{AdSwavefunction}
\psi(z) = N\sqrt{z}J_2(mz)
\end{align}
where $N$ is a normalization constant, and $J_2$ is a Bessel function of the first kind. (We dropped the $\sqrt{z}Y_2(mz)$ solution, which diverges at 0.)

Recalling that $\sqrt{z} J_2(z)\sim z^{5/4}$ as $z\rightarrow 0$ and $\sqrt{z}J_2(z)\sim \cos(z+3\pi/4)$, one can check that the wavefunction dies off near $z\rightarrow 0$, and is sinusoidal as $z\rightarrow \infty$, as expected for a scattering state. Here too there is no normalizable as solution, and the particle can have any 4-momentum (and any mass). 

Indeed, AdS$_5$ --  the original playground for holographic duality -- is {\em scale invariant}: invariant under dilations of the coordinates, or equivalently, under changing the energy scale. The only scale in the problem, $L$, cancelled out in the potential! The corresponding 4D theory must also look identical at different energy scales, which implies a continuous spectrum. 

To find a discrete spectrum, then, we must introduce a scale into the metric.

\subsection{The hard wall model}
The ``hard wall" model due to Erlich et al \cite{Erlich:2005qh} introduces a scale in AdS, creating an infinite square-well potential.  Its metric is identical to the AdS metric, except that the radial coordinate $z$ is cut off at a finite value, $z_0$:
\begin{align}
ds_{\rm{5d}}^2 = \frac{L^2}{z^2}\left( \eta_{\mu\nu}dx^\mu dx^\nu + dz^2 \right)~\qquad\rm{with}\qquad 0<z<z_0.
\end{align}
 $z_0$ is a free parameter, which one can fix by comparing the model's predictions for hadron masses and interactions to  experimental data. Since we focus only on the scaling of mass with excitation number, we  leave our results in terms of $z_0$.
 
\begin{figure}
     \centering
     \begin{subfigure}[b]{0.45\textwidth}
         \centering
         \includegraphics[width=\textwidth]{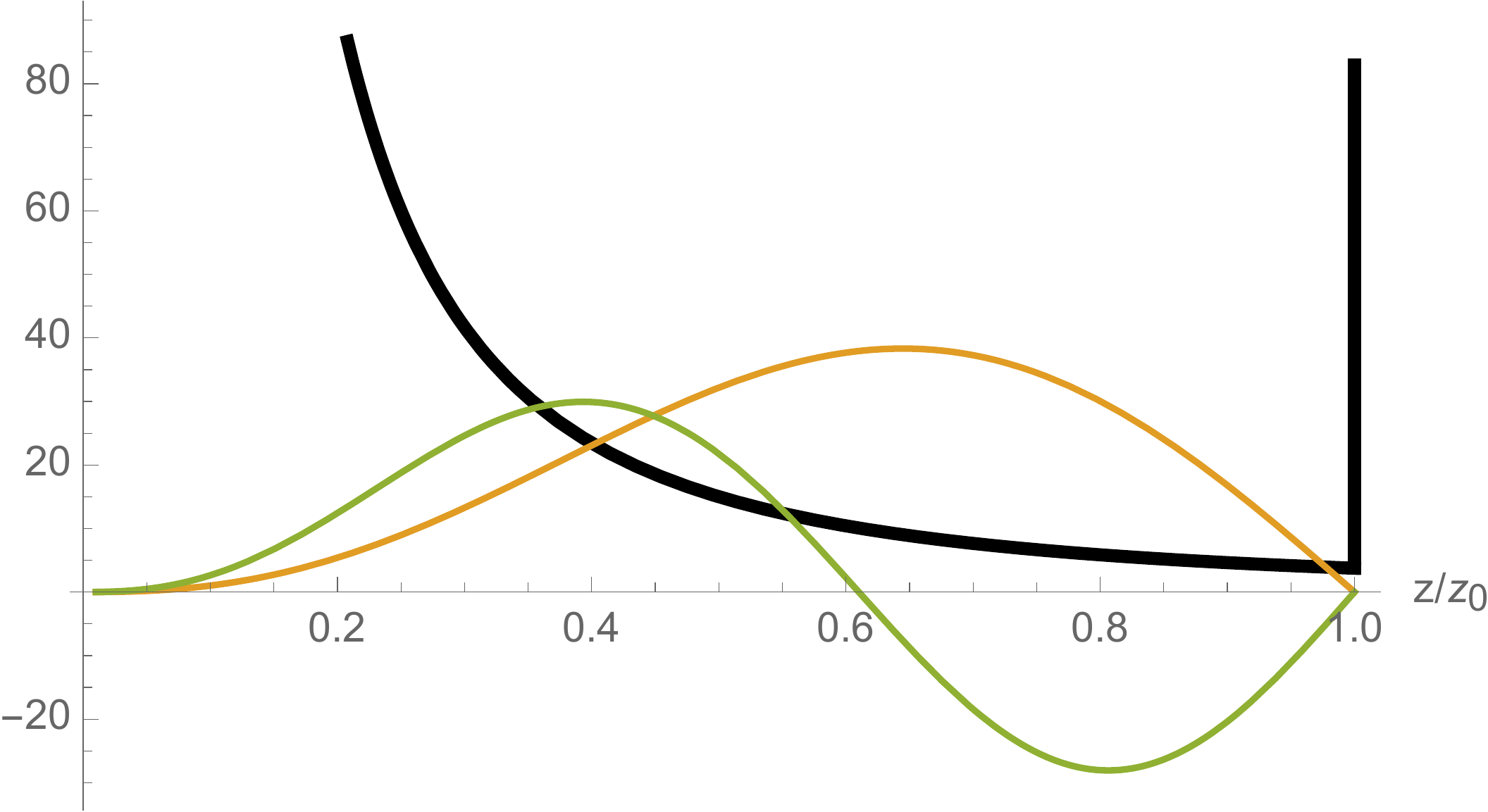}
         \caption{}
         \label{fig:HW_PotAndWFs}
     \end{subfigure}
     \hfill
     \begin{subfigure}[b]{0.35\textwidth}
         \centering
         \includegraphics[width=\textwidth]{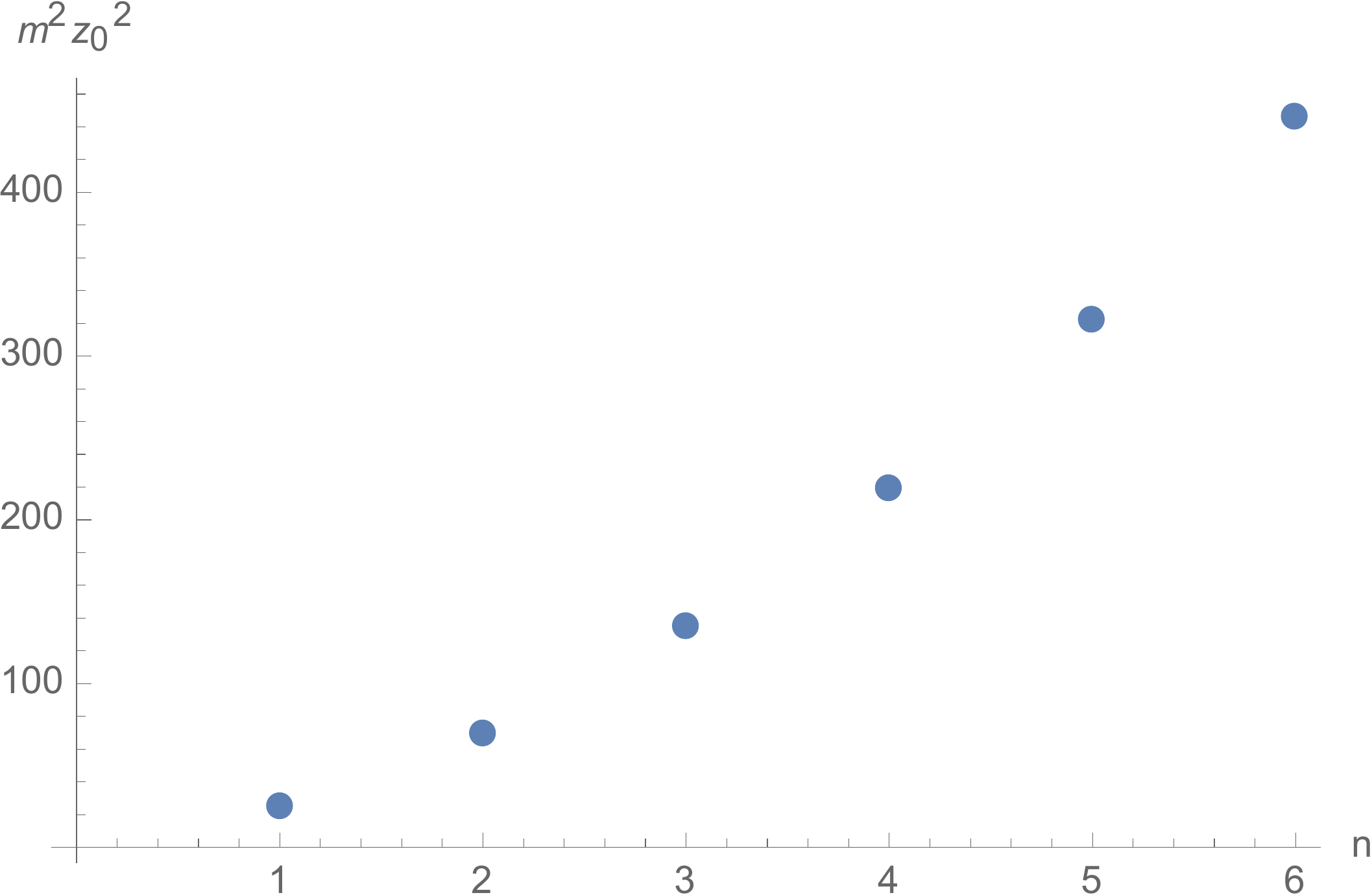}
         \caption{}
         \label{fig:HWmasses}
     \end{subfigure}
     \hfill
        \caption{The hard wall model. (a) The potential (black) and the first two wavefunctions (orange and green) as a function of the dimensionless coordinate $z/z_0$. (b) Predictions for (dimensionless) mass-squared $m^2z_0^2$ plotted as a function of excitation number $n$. Note that $m^2\sim n^2$ as for the quantum infinite square well.}\label{fig:HW}
\end{figure}

 The hard wall potential is identical to that of  AdS, except for the  wall at $z=z_0$ (see Figure \ref{fig:HW}). The allowed wavefunctions take the same form as well, except that the boundary condition imposing $\psi(z_0)=0$ at the wall restricts the possible values of the mass, as they do for the energy in the quantum mechanical infinite square well. The masses are  $m_n=\lambda_n/z_0$, where $\lambda_n=(5.1356.., 8.4172.., 11.6198.., ...)$ are zeroes of $J_2(mz)$. Each classical field $\Phi$ thus gives rise to a tower of hadron states.  The mass spacing scales like $1/z_0$, so  $z_0\rightarrow\infty$ indeed recovers the  continuous spectrum of pure AdS.

 As we can see in Figure \ref{fig:HW}, $m^2\sim n^2$. This is not surprising: the scaling behavior of the spectrum  is determined by the large $n$ behavior of the eigenvalues, or, equivalently, large arguments for the Bessel function $J_2(mz)$. For large $mz$, the Bessel functions are identical to the wave functions of the standard 1D square well: sines and cosines. Thus the eigenvalues of the two systems also have the same large $n$ scaling.

\subsection{The soft wall model}
Quantum mechanics provides a clue to a potential that {\em does} produce the correct, $m^2\sim n$ scaling for the spectrum: a harmonic oscillator. This was indeed the basis for the next iteration on hQCD, the ``soft wall model" \cite{Karch:2006pv}, which we derive a bit differently from the original formulation here.\footnote{While the soft wall model in its original form added an additional background field called the ``dilaton" on a pure AdS metric, the treatment we give here is equivalent. For practitioners: the difference amounts essentially to using Einstein frame instead of string frame.}

Consider a new  metric factor $\tilde{A}(z) \equiv \frac{L}{z} C(z)^{1/3}$, where $C(z)$ is a smooth function obeying $C(z\rightarrow 0)\sim 1$ to guarantee that the near-boundary behavior is AdS. The potential becomes
\begin{align}
V_{\text{sw}}(z) &=\frac{1}{4\tilde{A}^{6}}\left[ 2\tilde{A}^3\pd_z^2(\tilde{A}^3)-( \pd_z\tilde{A}^3 )^2\right]\cr
&=\frac{15}{4z^2}-\frac{1}{4}\left(\frac{C'}{C}\right)^2 - \frac{3}{2z}\frac{C'}{C}+\frac{C''}{2C} 
\end{align}
Once again, the scale $L$ does not appear in the potential. The first term is the usual repulsive piece from the AdS factor; the remaining terms should go like $z^2$ as $z\rightarrow\infty$ to get a harmonic oscillator.
Making the guess
\begin{align}
C(z) = e^{-(z/z_s)^\alpha}
\end{align}
for some arbitrary length scale $z_s$ and constant $\alpha$, 
\begin{align}
V_{\text{sw}}(z) - \frac{15}{4z^2}&=\frac{1}{4z^2}\left( \frac{z}{z_s}\right)^\alpha\left( 8\alpha -2 \alpha^2 +\alpha^2\left( \frac{z}{z_s}\right)^\alpha \right)~.
\end{align}
The final term dominates for $\alpha>0$, leading to the requirement $\alpha =2$. We now have
$\tilde{A}(z) = \frac{L}{z}e^{-(z/z_s)^2}$ and potential 
\begin{align}\label{VSW}
V_{\text{sw}}(z)&=\frac{1}{z_s^2}\bigg[ \frac{15}{4}\left(\frac{z}{z_s}\right)^{-2}+\left(\frac{z}{z_s}\right)^2 + 2\bigg]:=\frac{1}{z_s^2}\bigg[ \frac{15}{4}\frac{1}{\zt^2}+\zt^2 + 2\bigg]~,
\end{align}
where we introduce a dimensionless coordinate $\zt = \frac{z}{z_s}$ for convenience. 
 \begin{figure}
     \centering
     \begin{subfigure}[b]{0.45\textwidth}
         \centering
         \includegraphics[width=\textwidth]{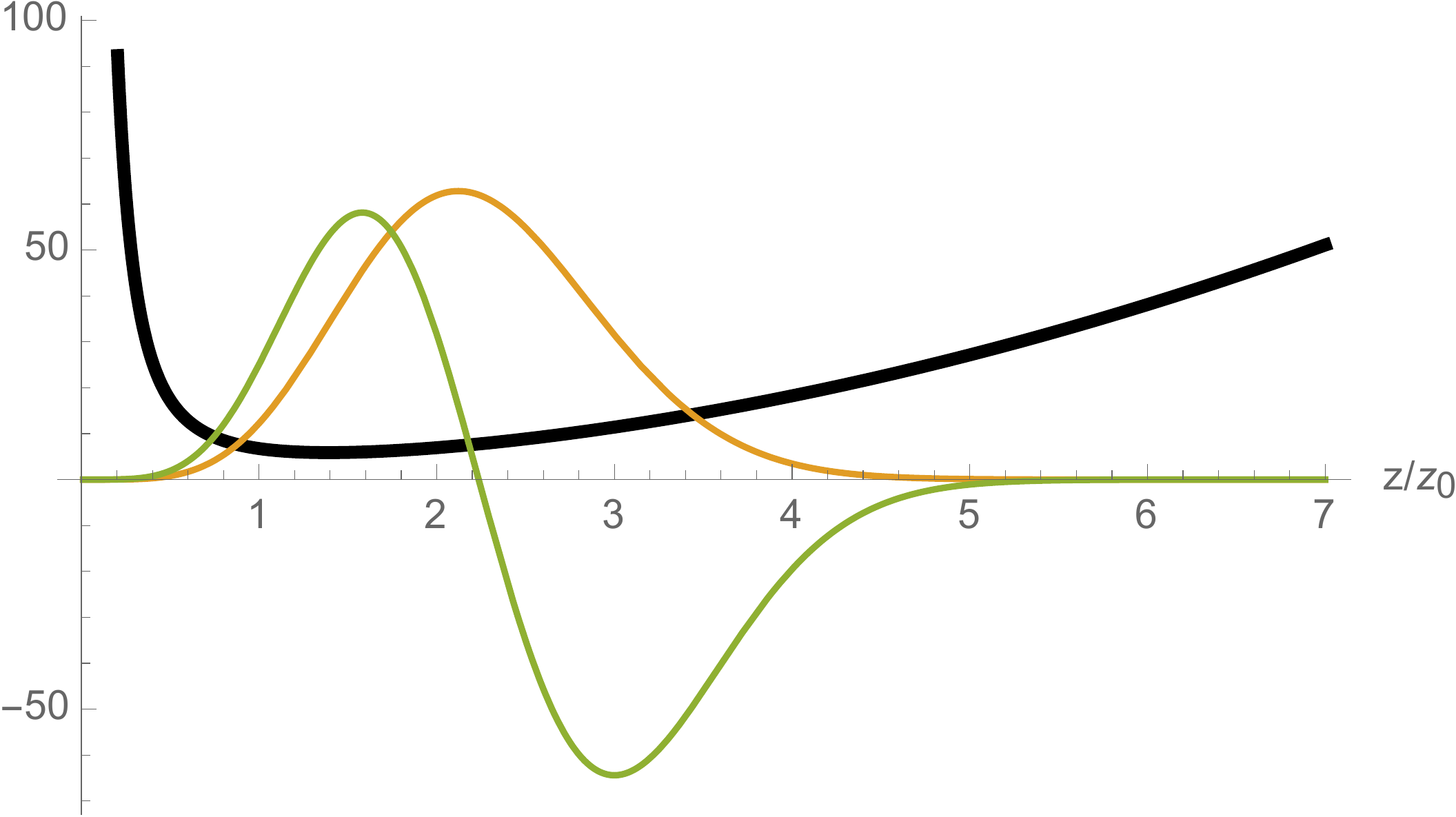}
         \caption{}
         \label{fig:SWpotential}         
     \end{subfigure}
     \hfill
     \begin{subfigure}[b]{0.35\textwidth}
         \centering
         \includegraphics[width=\textwidth]{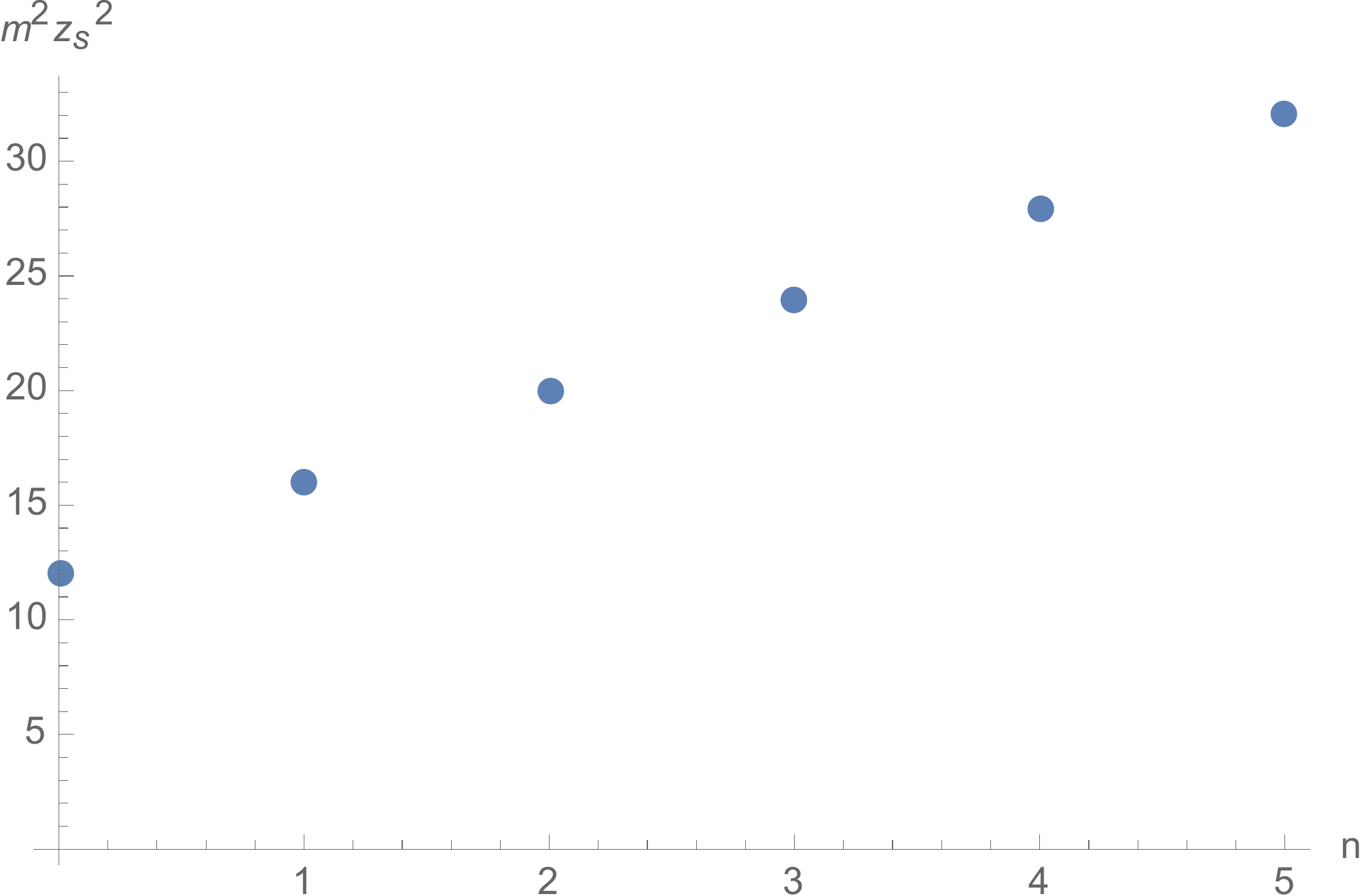}
         \caption{}
         \label{fig:SWmasses}
     \end{subfigure}
     \hfill
        \caption{The soft wall model. (a) The potential (black) and the first two wavefunctions (orange and green) as a function of the dimensionless coordinate $z/z_s$. (b) $m^2z_s^2$ plotted as a function of excitation number $n$.}
        \label{fig:SW}
\end{figure}

We can solve for the wavefunctions analytically as sketched in \cite{Karch:2006pv} using standard techniques from the higher-dimensional quantum harmonic oscillator problem. After absorbing the constant term of the potential \eqref{VSW} into the eigenvalue as $\lambda = z^{2}_{s}m^{2} - 2$, and changing variables to $\zt$, the Schr\"{o}dinger equation takes the  form
\begin{equation}
\psi^{''} - \left( \zt^{2} + \frac{k^{2} - \frac{1}{4}}{\zt^{2}} \right)\psi =  -\lambda \psi
\end{equation}
where $k=4$. Noting that $k^{2} - \frac{1}{4} = (k + \frac{1}{2})(k - \frac{1}{2})$, we guess a solution of the form
\begin{equation}
\psi \propto  \zt^{k + \frac{1}{2}}g(\zt)e^{-\frac{\zt^{2}}{2}}    
\end{equation}
where $g(\zt)$ is a polynomial in $\zt$. This behaves similarly to the standard quantum harmonic oscillator wavefunctions at infinity, but also has the modification $\zt^{k + \frac{1}{2}}$ due to the $\zt^{-2}$ term in the potential. The $\zt^{k+\frac{1}{2}}$ term guarantees that the wavefunction will vanish at the origin. Furthermore, when we take the second derivative of the wavefunction, the term $(k + \frac{1}{2})(k - \frac{1}{2}) \zt^{k - \frac{3}{2}}g(\zt)e^{-\frac{\zt^{2}}{2}} $ cancels the $\zt^{-2}$  in the potential. With this ansatz, and making the substitution $u = \zt^{2}$, we obtain
\begin{equation}
ug^{''} + (k + 1 - u)g^{'} - \left(\frac{k+1}{2} - \frac{\lambda}{4}\right)g = 0.
\end{equation}
This is the second order differential equation corresponding to the associated Laguerre polynomials: 
\begin{equation}
ug^{''} + (k + 1 - u)g^{'} + (n-1)g = 0
\end{equation}
where here $n$ is a positive integer. Equating the coefficients for $g$ we obtain the eigenvalues
\begin{equation}
\lambda = 4n +2k + 2 \quad \Rightarrow \quad m^{2} = \frac{1}{z_{s}^{2}}(4n + 2k + 4)
\end{equation}
which correspond to the associated Laguerre polynomials $L_{n-1}^{k}(u) = L_{n-1}^{k}(\zt^{2})$. As expected, the eigenvalues are linear in $n$, and the constant term in the potential gives rise to a linear shift in the mass-squared, as it would for the energies of any quantum mechanical potential. The normalized eigenfunctions for our case,  $k=4$, take the form
\begin{equation}
\psi_{n} =  \zt^{\frac{9}{2}} \sqrt{ \frac{2(n-1)!}{(n+3)!}} L_{n-1}^{4}(\zt^{2})e^{-\frac{\zt^{2}}{2}}\quad \text{where}\quad n=1,2,3,\dots.
\end{equation}
The first two of these, together with the potential, are shown in Figure \ref{fig:SW}.

\section{Conclusions and Further Elaborations}

We have shown that the problem of choosing an appropriate 5D metric for hQCD models reduces to choosing a potential in a 1D Schr\"{o}dinger equation. We then illustrated the method on examples from the literature with direct analogs to well-known quantum-mechanics problems (the free particle, infinite square well, and harmonic oscillator).

One could consider many elaborations on and generalizations of this technique. For instance:
\begin{enumerate}
\item While our focus was hQCD, the method described could give students who have not studied General Relativity intuition for the meaning of curved space. Minima of the Schr\"{o}dinger potential for a given metric (or, equivalently, maxima of the ground state wavefunction) correspond to locations in spacetime where particles' energies are minimized. In the case of the asymptotically AdS spacetimes studied here, for instance, the metric pushes particles away from the boundary at $z=0$. One can apply the same method to other spacetimes (like de Sitter, Schwarzschild, etc.).
\item We focused on scalar fields in the 5D that are dual to scalar hadrons in 4D. One can also study higher spin classical fields, dual to higher spin hadrons \cite{Karch:2006pv}. The essential difference in the form of the Schr\"{o}dinger potential comes from additional factors of the  metric that appear in the higher spin fields' action. For example, vector hadrons (like the $\rho$ meson) correspond to massless vector fields in 5D. Their 5D kinetic terms go like $\pd_M A_N \pd_P A_Q g^{MP}g^{NQ}$, where the contraction of the vector field's index yields a extra factor of $A(z)^{-2}$.
\item The hard wall model \cite{Erlich:2005qh} differs from the infinite square well of quantum mechanics because it allows boundary conditions at $z=z_0$ other than $\psi(z_0)=0$ (since the finite size of the $z$ interval still allows the solution to be normalizable). In fact, maintaining certain symmetries of QCD -- like isospin -- requires boundary conditions like $\varphi'(z_0)=0$ for certain fields. One can therefore explore the effect of different sets of boundary conditions (Neumann, Dirichlet, or mixed) on the spectrum.
\end{enumerate}

{\bf Acknowledgements:} SKD's work is supported by NSF grant PHY-2014025 and by a New York Tech ISRC grant. TL's and PM's summer and fall internships with SKD were supported by the New York Tech ISRC. SKD thanks Nelia Mann for extremely helpful comments on this manuscript.


\end{document}